\documentclass{amia}
\usepackage{graphicx}
\usepackage[labelfont=bf]{caption}
\usepackage[superscript,nomove]{cite}
\usepackage{color}
\usepackage{wrapfig,lipsum,booktabs}

\begin{document}

\title{Machine Learning for removing EEG artifacts: Setting the benchmark}

\author{Subhrajit Roy, PhD}

\institutes{
    IBM Research Australia\\
}

\maketitle

\noindent{\bf Abstract}
\textit{Electroencephalograms (EEG) are often contaminated by artifacts which make interpreting them more challenging for clinicians. Hence, automated artifact recognition systems have the potential to aid the clinical workflow. In this abstract, we share the first results on applying various machine learning algorithms to the recently released world's largest open-source artifact recognition dataset. We envision that these results will serve as a benchmark for researchers who might work with this dataset in future.}

\section*{Introduction}
Removal of artifacts from electroencephalogram (EEG) is a necessary step in analyzing EEG signals since artifacts can lead to severe misinterpretation of these signals. However, manual removal of artifacts requires trained clinicians or neurophysiologists and is a procedure that is known to be both time and resource hungry. This is due to the large amount of data recorded and the availability of limited number of certified investigators. In contrast, automated identification and removal of artifacts offer the potential to improve the EEG analysis process by reducing the analysis time. Moreover, this can potentially improve the quality of patient care by shortening the time to diagnosis and reducing manual error. Hence, in literature, the design of automated systems for artifact removal have received considerable research focus \cite{ref1,ref2,ref3}. However, most of these studies suffer from either of or both the following problems: the dataset used is closed-source thereby making reproduction of the results impossible or it is too small for a statistically significant conclusion. In our work, we try to address this issue by using the recently open-sourced NEDC TUH EEG Artifact Corpus \cite{ref4} containing 66638 seconds of artifact data. Since this is a new dataset, there has been no benchmark results reported yet. In this abstract, we address this gap by sharing the performance of various machine learning algorithms on the artifact recognition task based on this dataset. We hope that the results reported in this abstract will serve as a benchmark for researchers and developers working with this dataset in future.


\section*{Methods}
\begin{wraptable}{r}{0.52\textwidth}
	\caption{Description of NEDC TUH EEG Artifact Corpus.}\label{wrap-tab:1}
	\begin{tabular}{cccc}\toprule  
		Artifact type & \# patients & \# sessions & \# seconds \\\midrule \midrule
		Eye movements & 140 & 166 & 24064\\  \midrule
		Chewing & 22 & 23 & 10646\\  \midrule
		Shivering & 14 & 14 & 4005\\  \midrule
		Electrode pops & 80 & 97 & 9057\\  \midrule
		Muscle movements & 74 & 90 & 18866\\  \midrule
		Null & 213 & 259 & 1366299\\  \bottomrule
	\end{tabular}
\end{wraptable} 
The NEDC TUH EEG Artifact Corpus consists 259 EEG sessions collected from 213 patients. It has the following five types of artifact events: eye movements, chewing, shivering, electrode pops, and muscle movements. Moreover, for each session, any segment which is not an artifact is marked as null class. This is included since the goal is to automatically annotate and remove artifacts from EEG sessions that might contain various types of events. This makes the problem a six-class classification task. Moreover, to successfully remove an artifact or parts of an artifact it is necessary to identify as much of it as possible. Hence, instead of taking event-based decisions, our goals is to annotate every unit (1 second) of time. The dataset was segregated at the patient level into training (60\%), validation (20\%), and test (20\%) sets. This is to enforce that the algorithms can generalize across patients. We pre-processed each set by converting the recorded raw EEG signal into the transverse central parietal (TCP) montage system for accentuating spike activity leading to a 22-channel output. Next, we applied a popular technique used for extracting features from EEG signals \cite{ref5}. We computed FFT for each second of clip (with 75\% overlap) across all EEG channels. Next, the output of FFT is clipped from 1 to 24 Hz and normalized across frequency buckets and the correlation coefficients $(22,22)$ matrix is calculated from this normalized matrix. We computed the absolute values of the eigenvalues of the upper right traingle of the correlation coefficients matrix and fed them to the machine learning algorithms. For classification, we chose one popular algorithm from each major family of supervised learning algorithms: AdaBoost, Gaussian Naive Bayes (GaussianNB), k-Nearest Neighbors (k-NN), Linear Discriminant Analysis (LDA), Multi-layer Perceptrons (MLP), Random Forests, SGD classifier, and XGBoost. For each of these algorithms, we used the HyperOpt \cite{ref6} meta-algorithm to explore and find the optimal hyperparameters based on its performance on the validation set. Moreover, since the dataset is imbalanced, during training these classifiers, we under-sample the majority classes by randomly picking samples without replacement.

\section*{Results and discussion}
The performance of different algorithms explored in this paper is shown in Table \ref{wrap-tab:1}.  For each algorithm, we report the weighted-F1 score (since the dataset is imbalanced), accuracy, and sensitivities of recognizing each second of an artifact and the null class of the test set. $S_{eyem}$, $S_{chew}$, $S_{shiv}$, $S_{elpp}$, and $S_{musc}$ denote the sensitivities in identifying EEG artifacts due eye movements, chewing, shivering, electrode pops, and muscle movements respectively. $S_{null}$ denotes the sensitivity in correctly identifying non-artifacts. Table \ref{wrap-tab:1} depicts that LDA achieves the best overall weighted-F1 score and accuracy. However, it doesn't produce the best performance for recognizing all artifacts. Moreover, we observe that the classifiers are being unsuccessful in identifying EEG artifacts produced due to shivering. 

\begin{table}
	\centering
	\caption{The performance (averaged over 5 runs) of various algorithms for artifact recognition. }\label{wrap-tab:1}
	\begin{tabular}{ccccccccc}\toprule  
		Algorithm & Weighted-F1 score & Accuracy & $S_{eyem}$ & $S_{chew}$ & $S_{shiv}$ & $S_{elpp}$ & $S_{musc}$ & $S_{null}$ \\\midrule \midrule
		 AdaBoost & 0.7375 & 62.57\% & 62.51\% & 68.63\% & 2.31\% & 28.30\% & 62.88\% & 63.17\%  \\\midrule 
		 GaussianNB & 0.7773 & 67.79\% & 63.19\% & 72.67\% & \textbf{16.32}\% & 13.99\% & 43.47\% & 69.03\% \\\midrule 
		 k-NN & 0.7476 & 63.76\% & 60.77\% & \textbf{86.07}\% & 5.95\% & 26.06\% & 48.55\%  & 64.67\% \\\midrule 
		 LDA & \textbf{0.8012} & \textbf{71.43}\% & 58.73\% & 62.73\% & 2.50\% & 26.99\% & \textbf{70.76}\% & \textbf{72.39}\%  \\\midrule
		 MLP & 0.7787 & 68.23\% & 68.36\% & 73.80\% & 2.05\% & 34.64\% & 65.35\% & 68.93\% \\\midrule
		 Random Forests & 0.7834 & 68.80\% & \textbf{73.35}\% & 80.35\% & 3.00\% & 35.26\% & 67.25\%  &69.39\%\\\midrule 
		 SGD classifier & 0.7887 & 69.57\% & 63.06\% & 73.61\% & 3.10\% & 28.79\% & 69.01\%  &70.36\%\\\midrule 
		 XGBoost & 0.7996 & 71.19\% & 72.38\% & 74.08\% & 2.75\% & \textbf{38.75}\% & 67.91\%  &71.90\%\\\midrule
	\end{tabular}
\end{table} 

\section*{Conclusion}
In this paper, we evaluate the performance of various machine learning algorithms for artifact recognition on the world's largest open-source EEG artifacts dataset. The dataset was publicly released on December 2018 and to the best of our knowledge this is the first paper to share benchmark results. We find that while machine learning can obtain promising results on this dataset, there is room for improvement particularly in identification of EEG artifacts due to shivering. We envision that the results reported in this abstract will serve as a benchmark for this important task of automated artifact recognition and removal.

\makeatletter
\renewcommand{\@biblabel}[1]{\hfill #1.}
\makeatother

\bibliographystyle{unsrt}

\begin{thebibliography}{1}
\setlength\itemsep{-0.1em}

\bibitem{ref1}
Stone DB, Gabriella T, Patrique F, Jens H, Silvia C. Automatic Removal of Physiological Artifacts in EEG: The Optimized Fingerprint Method for Sports Science Applications. Frontiers in Human Neuroscience . 2018;12:96.
\bibitem{ref2}
Raduntz T, Scouten J, Hochmuth O, Meffert B.
EEG artifact elimination by extraction of ICA-component features using image processing algorithms.
Journal of Neuroscience Methods. 2015;243:84-93.
\bibitem{ref3}
Kang G, Jin SH, Kim DK, Kang SW. EEG artifacts removal using machine learning algorithms and independent component analysis. Clinical Neurophysiology. 2018;129:e24.
\bibitem{ref4} 
Dataset Link. Temple university EEG corpus. Link: https://www.isip.piconepress.com/projects/tuh\_eeg/.
\bibitem{ref5}
Schindler K, Leung H, Elger CE, Lehnertz K. Assessing seizure dynamics by analysing the correlation structure of
multichannel intracranial EEG. Brain. 2007;130:65–77.
\bibitem{ref6} Bergstra J. Hyperopt: a Python library for model selection and hyperparameter optimization. Computational Science \& Discovery. 2015;8:1.





\end{thebibliography}

\end{document}